\documentclass[journal]{IEEEtran}
\usepackage{multicol}
\usepackage{graphicx}
\usepackage[latin1]{inputenc}
\usepackage{amsmath}
\usepackage{amsfonts}
\usepackage{acronym}
\usepackage{amsthm}

\usepackage{CJK}
\usepackage{algorithm}

\usepackage{amssymb}
\usepackage{dsfont}    
\usepackage{epsfig}
\usepackage{ifthen}
\usepackage{psfrag}
\usepackage{url}
\usepackage{listings}
\usepackage{comment}
\usepackage{fancyhdr}
\usepackage[newitem,newenum]{paralist}
\usepackage{booktabs}
\usepackage{tabularx}
\usepackage{color}
\usepackage{cite}
\usepackage{morefloats}
\usepackage{setspace}
\usepackage[newitem,newenum]{paralist}
\usepackage{booktabs}
\usepackage[normalsize]{subfigure}
\usepackage{epstopdf}
\usepackage{algorithmic,algorithm}

\usepackage{color}

\begin{document}
%
\title{QoS-Based Source and Relay Secure Optimization Design with Presence of Channel Uncertainty}
\author{Meng~Zhang, Jian~Huang, Hui~Yu, Hanwen~Luo and Wen~Chen, \textit{Senior Member}, \textit{IEEE}

\thanks{The authors are with the Dept. of Electronic Engineering, Shanghai Jiao Tong Univ., P. R. China (email:\{mengzhang, 1250603hj, yuhui, hwluo, wenchen\}@sjtu.edu.cn).}

}
\maketitle


\begin{abstract}
In this letter, we study relay-aided networks with presence of single eavesdropper. We provide joint beamforming design of the source and relay that can minimize the overall power consumption while satisfying our predefined quality-of-service (QoS) requirements. Additionally, we investigate the case that the channel between relay and eavesdropper suffers from channel uncertainty. Finally, simulation results are provided to verify the effectiveness of our algorithm.
\end{abstract}


\setlength\arraycolsep{2pt}
\section{Introduction}
Recently, research concerning secrecy capacity has captured considerable attentions, though initial concept of secure communication
can be dated back to the 1970s~\cite{paper:1}. Traditional high layer encryption-based method can hardly be applied in
certain circumstances, e.g., wireless local area network (WLAN) or Ad hoc networks. Due to the fact that users' random accessing and leaving are difficult
to predict in WLAN scenario, establishing an appropriate high layer protocol is not an easy task. Additionally,  in Ad hoc networks a complete transmission might take several hops and be relayed by other users. Consequently, how to guarantee secure communication has become a critical issue.

Roughly speaking, the research in this area can be classified into three categories. The first category falls into the artificial-noise based algorithm that relies
on generating additional noise bringing more negative effect to the eavesdropper than to the legal user. In~\cite{paper:2}, the authors investigate a point-to-point system
with the presence of an eavesdropper and it has been shown how secrecy can be achieved by adding artificial noise. The second
 category falls into beamforming based algorithm. For instance, a joint beamforming design of relay and source is proposed in~\cite{paper:3} with the
assumption that the relay also plays as an eavesdropper that tends to wiretap the user's message. The last category is a combination of the above two sorts.
 Specifically, in~\cite{paper:4} the authors study a broadcast scenario by utilizing both the artificial noise and beamforming together and simulation results demonstrate that
 joint design can achieve better performance.

It should be noticed that all the above studies are based on the perfect channel state information (CSI) assumptions.  Although the channel between
 relay and legal user can be obtained through uplink feedback, such assumption is not appropriate for the channel between eavesdropper and relay since
eavesdropper usually behaves in passive manner. Therefore, it is more practical to consider the imperfect CSI cases. In~\cite{paper:5}, the
authors investigate a multipoint-to-mutlipoint system under norm-bounded error model and propose precoding designs that can maximize the users' signal-to-interference-plus-noise-ratio (SINR). Besides, relay-aided  multiple source-destination pairs networks have been studied in~\cite{paper:6}, where all channels suffer from norm-bounded errors. The authors provide relay precoding strategy that can minimize the power consumption while maintaining certain quality-of-service (QoS) requirements. Moreover, in~\cite{paper:7} the authors tend to maximize the legal user's SINR while constraining the eavesdropper's SINR below a threshold.

In this letter, we will
study relay-aided networks that beamforming technology is adopted at both source and relay. Additionally, we assume that the channel between relay and
eavesdropper is not perfect, specifically, following the norm-bounded model. Our target is to minimize the sum power consumption of
relay and source while satisfying the legal user's QoS requirement and maintaining the eavesdropper's signal-to-noise-ratio (SNR) below a threshold.

$\textit{Notations}$: In this paper, we use bold uppercase and lowercase letters denote matrices and vectors, respectively; $(\cdot)^{*}$,$(\cdot)^{T}$ and $(\cdot)^{H}$ to denote the conjugate, transpose and conjugate transpose of a matrix or a vector, respectively; $\mathbf{I}_{N}$ is an $N\times{N}$ identity matrix; $\mathbb{E}{(\cdot)}$ denotes the statistical expectation; $Tr(\cdot)$  and $\mathfrak{Re}{\{}\cdot{\}}$ are the trace of a matrix and the real part of a variable, respectively; $vec(\cdot)$ represents the matrix vectorization; $\otimes$ denotes the Kronecker product; $\parallel{\cdot}\parallel$ denotes the Frobenius norm; $\succeq$ represents the property of semidefinite.

\section{System Model}
 Throughout this letter, we assume that Bob, equipped with single antenna, is a legal subscriber of cellular networks. At the same time, there also exists a single-antenna eavesdropper wiretapping the transmitting data for Bob.
Besides, it is supposed that direct communication between source and Bob is inapplicable mainly due to the large-scale fading caused by long distance between them. As a result, relay technology has to be introduced so as to help the transmission shown in Fig.~1. The source and relay are equipped with $N$ antennas and $M$ antennas, respectively. Moreover, two timeslots are needed to complete a transmission process.
In the first timeslot, the source transmits the message intended for Bob, which can be expressed as
\begin{eqnarray}\label{eq:1}
\mathbf{s}=\mathbf{q}x,
\end{eqnarray}
where $\mathbf{q}\in\mathbb{C}^{N\times{1}}$ denotes the beamforming vector executed at source; $x$ is the intended data for Bob which satisfies $\mathbb{E}\{xx^{*}\}=1$. The signal received at relay can be written as
\begin{eqnarray}\label{eq:2}
\mathbf{y}_r=\mathbf{H}\mathbf{s}+\mathbf{n}_r,
\end{eqnarray}
where $\mathbf{H}\in\mathbb{C}^{M\times{N}}$ represents the channel between relay and source; $\mathbf{n}_r\in\mathbb{C}^{M\times{1}}$ is the additive Gaussian noise which satisfies $\mathbb{E}\{\mathbf{n}_r\mathbf{n}_r^{H}\}=\sigma_r^2\mathbf{I}_M$. Afterwards, the received data at relay will be multiplied by precoding matrix $\mathbf{W}\in\mathbb{C}^{M\times{M}}$,
\begin{eqnarray}\label{eq:3}
\mathbf{x}_r=\mathbf{W}\mathbf{y}_r=\mathbf{W}\mathbf{H}\mathbf{q}x+\mathbf{W}\mathbf{n}_r.
\end{eqnarray}

In the second timeslot, relay will broadcast the signal $\mathbf{x}_r$. The received signal at Bob can be expressed as
\begin{eqnarray}\label{eq:4}
{y}_b=\mathbf{g}_b\mathbf{x}_r+{n}_b=\mathbf{g}_b\mathbf{W}\mathbf{H}\mathbf{q}x+\mathbf{g}_b\mathbf{W}\mathbf{n}_r+{n}_b,\quad\,
\end{eqnarray}
where $\mathbf{g}_b\in\mathbb{C}^{1\times{M}}$ is the channel between relay and Bob which can be acquired by the feedback information from Bob and ${n}_b$ is the additive Gaussian noise at Bob satisfying $\mathbb{E}\{{n}_b{n}_b^{*}\}=\sigma_b^2$. In this letter, we assume that the channel knowledge about $\mathbf{g}_b$ is perfect.
However, the channel between relay and eavesdropper cannot be guaranteed to be perfect. In this letter, we will adopt a norm-bound error model where the norm of channel estimation error is inferior to a threshold. The channel between relay and eavesdropper can be presented as
\begin{eqnarray}\label{eq:5}
\mathbf{g}_e=\mathbf{\bar{g}}_e+\triangle\mathbf{g}_e,\|\triangle\mathbf{g}_e\|\leq{\varepsilon},
\end{eqnarray}
where $\mathbf{\bar{g}}_e\in\mathbb{C}^{1\times{M}}$ is estimated channel between eavesdropper and relay and $\triangle\mathbf{g}_e\in\mathbb{C}^{1\times{M}}$ is the channel estimation errors bounded by radius $\varepsilon$.
Similarly, the received signal at eavesdropper is expressed as
\begin{eqnarray}\label{eq:6}
{y}_e=\mathbf{g}_e\mathbf{x}_r+{n}_e=\mathbf{g}_e\mathbf{W}\mathbf{H}\mathbf{q}x+\mathbf{g}_e\mathbf{W}\mathbf{n}_r+{n}_e,\quad\
\end{eqnarray}
where ${n}_e$ is additive Gaussian noise satisfying $\mathbb{E}\{{n}_e{n}_e^{*}\}=\sigma_e^2$.
\begin{figure}[!htbp]
\centering
  \includegraphics[width=8cm,height=3cm]{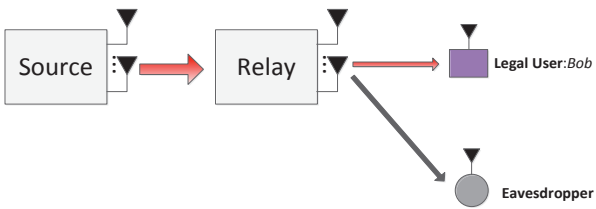}
\caption{Relay-aided networks with presence of single eavesdropper}
\label{fig_network_model}
\end{figure}
\section{Joint Source and Relay Beamforming Design with Presence of Channel Uncertainty}
In this letter, we aim to minimize the entire power consumption at source and relay while satisfying predefined QoS requirement for Bob and simultaneously constraining the SNR of eavesdropper below certain threshold, respectively. The SNR of Bob and eavesdropper can be expressed as
\begin{small}
\begin{eqnarray}\label{eq:7}
{\textmd{SNR}}_b&=&\frac{\mathbf{g}_b\mathbf{W}\mathbf{H}\mathbf{q}\mathbf{q}^{H}\mathbf{H}^{H}\mathbf{W}^{H}\mathbf{g}_b^{H}}
{\sigma_r^{2}\mathbf{g}_b\mathbf{W}\mathbf{W}^{H}\mathbf{g}_b^{H}
+\sigma_b^2},
\end{eqnarray}
and
\begin{eqnarray}\label{eq:8}
{\textmd{SNR}}_e&=&\frac{\mathbf{g}_e\mathbf{W}\mathbf{H}\mathbf{q}\mathbf{q}^{H}\mathbf{H}^{H}\mathbf{W}^{H}\mathbf{g}_e^{H}}
{\sigma_r^{2}\mathbf{g}_e\mathbf{W}\mathbf{W}^{H}\mathbf{g}_e^{H}
+\sigma_e^2}.
\end{eqnarray}
\end{small}
Hence, our optimization problem can be formulated as
\begin{subequations}\label{eq:9}
\begin{eqnarray}
 &&\mathop {\textrm{min}}_{\mathbf{q},\mathbf{W}}\qquad{\mathbb{E}(\|\mathbf{s}\|^{2})+\mathbb{E}(\|\mathbf{x}_r\|^{2})},\\
 &&s.t.\quad \quad{\textmd{SNR}}_b\geq{r_{th}^{(b)}},\\
  &&\qquad\quad\;{\textmd{SNR}}_e\leq{r_{th}^{(e)}},\|\triangle\mathbf{g}_e\|\leq{\varepsilon},
 \end{eqnarray}
 \end{subequations}
 where $r_{th}^{(b)}$ and $r_{th}^{(e)}$ denote the predefined thresholds for Bob and eavesdropper, respectively.

Define $\mathbf{Q}=\mathbf{q}\mathbf{q}^{H}$, the base station power can be turned into\
\begin{eqnarray}\label{eq:10}
\mathbb{E}(\|\mathbf{s}\|^{2})=Tr(\mathbf{Q}).
\end{eqnarray}
By introducing $\mathbf{w}=vec(\mathbf{W})$ and $\mathbf{Z}=\mathbf{w}\mathbf{w}^{H}$, and with the help of equalities $Tr(\mathbf{X}\mathbf{Y}\mathbf{X}^{H}\mathbf{W})=vec(\mathbf{X})^{H}(\mathbf{W}^{T}\otimes\mathbf{Y})vec(\mathbf{X})$ and $Tr(\mathbf{A}\mathbf{B})=Tr(\mathbf{B}\mathbf{A})$~\cite{paper:8}, the relay's power can be transformed as
\begin{eqnarray}\label{eq:11}
\nonumber
&&\mathbb{E}(\|\mathbf{x}_r\|^{2})\\\nonumber
&=&Tr\big(\mathbf{W}\mathbf{H}\mathbf{q}\mathbf{q}^{H}\mathbf{H}^{H}\mathbf{W}^{H}+\sigma_r^{2}\mathbf{W}\mathbf{W}^{H}\big)\\\nonumber
&=&Tr\big(\mathbf{w}^{H}\big(\mathbf{I}_M\otimes\big(\mathbf{H}\mathbf{q}\mathbf{q}^{H}\mathbf{H}^{H}+\sigma_r^{2}\mathbf{I}_M\big)\big)\mathbf{w}\big)\\
&=&Tr\big(\mathbf{Z}\big(\mathbf{I}_M\otimes\big(\mathbf{H}\mathbf{Q}\mathbf{H}^{H}+\sigma_r^{2}\mathbf{I}_M\big)\big)\big).
\end{eqnarray}
Similarly, SNR of Bob can be rewritten as
\begin{eqnarray}\label{eq:12}
\nonumber
{\textmd{SNR}}_b&=&\frac{\mathbf{w}^{H}\big((\mathbf{g}_b^{H}\mathbf{g}_b)^{T}\otimes(\mathbf{H}\mathbf{Q}\mathbf{H}^{H})\big)\mathbf{w}}
{\mathbf{w}^{H}\big((\mathbf{g}_b^{H}\mathbf{g}_b)^{T}\otimes(\sigma_r^{2}\mathbf{I}_M)\big)\mathbf{w}
+\sigma_b^2}\\
&=&\frac{Tr\big(\mathbf{Z}\big((\mathbf{g}_b^{H}\mathbf{g}_b)^{T}\otimes(\mathbf{H}\mathbf{Q}\mathbf{H}^{H})\big)\big)}
{Tr\big(\mathbf{Z}\big((\mathbf{g}_b^{H}\mathbf{g}_b)^{T}\otimes(\sigma_r^{2}\mathbf{I}_M)\big)\big)
+\sigma_b^2}.
\end{eqnarray}
Nevertheless, the SNR of eavesdropper is difficult to handle due to the presence of channel uncertainty. Therefore, we resort to optimizing the worst case of eavesdropper's SNR. Here, we will separately find the upper bound of the numerator of ${\textmd{SNR}}_e$ and lower bound of the denominator of ${\textmd{SNR}}_e$, respectively.

Before explicit computations of the lower and upper bounds, we will state the following two useful results~\cite{paper:6} that will be utilized later. For the following two problems
\begin{eqnarray}\label{eq:13}
     &&\mathop{\textrm{max}}_{\|\mathbf{x}\|\leq{\delta}}{\quad}\mathcal{X}(\mathbf{x})=\mathfrak{Re}(\mathbf{x}^{H}\mathbf{y}),\\
     &&\mathop{\textrm{min}}_{\|\mathbf{x}\|\leq{\delta}}{\quad}\mathcal{Y}(\mathbf{x})=\mathfrak{Re}(\mathbf{x}^{H}\mathbf{y}),
\end{eqnarray}
their solutions can be given by
\begin{eqnarray}
\label{eq:14a}&&\mathcal{X}(({\delta}/\|\mathbf{y}\|)\mathbf{y})\,\;\;={\delta}\|\mathbf{y}\|,\\
\label{eq:14b}&&\mathcal{Y}(-({\delta}/\|\mathbf{y}\|)\mathbf{y})\,=-{\delta}\|\mathbf{y}\|.
\end{eqnarray}

Then, given $\mathbf{X}_{1}\in{\mathbb{C}}^{N_{1}{\times}N_{2}}$, $\mathbf{F}\in{\mathbb{C}}^{N_{2}{\times}N_{3}}$, $\mathbf{X}_{2}\in{\mathbb{C}}^{N_{3}{\times}N_{3}}$ and $\mathbf{X}_{3}\in{\mathbb{C}}^{N_{2}{\times}N_{4}}$, the following equalities hold
\begin{eqnarray}\label{eq:15}
 \nonumber
 &&\quad\|\mathbf{X}_{1}\mathbf{F}\mathbf{X}_{2}\mathbf{F}^{H}\mathbf{X}_{3}\|\\\nonumber
 &&\mathop{=}^{(a)}\left\|vec(\mathbf{X}_{1}\mathbf{F}\mathbf{X}_{2}\mathbf{F}^{H}\mathbf{X}_{3})\right\|\\\nonumber
 &&\mathop{=}^{(b)}\left\|(\mathbf{X}_{3}^{T}\otimes\mathbf{X}_{1})vec(\mathbf{F}\mathbf{X}_{2}\mathbf{F}^{H})\right\|\\\nonumber
 &&\mathop{=}^{(c)}\left\|(\mathbf{X}_{3}^{T}\otimes\mathbf{X}_{1})(\mathbf{F}^{*}\otimes\mathbf{F})vec(\mathbf{X}_{2})\right\|\\
 &&\mathop{=}^{(d)}\left\|\left(vec(\mathbf{X}_{2})^{T}\otimes(\mathbf{X}_{3}^{T}\otimes\mathbf{X}_{1})\right)vec(\mathbf{F}^{*}\otimes\mathbf{F})\right\|,
\end{eqnarray}
where the equality (a) holds with the help of the equation $\|\mathbf{X}\|=\|vec(\mathbf{X})\|$; the equalities (b) (c) and (d) hold with the help of $vec(\mathbf{ABC})=(\mathbf{C}^{T}{\otimes}\mathbf{A})vec(\mathbf{B})$~\cite{paper:8}.

Furthermore, we define $\mathbf{f}=vec(\mathbf{F})$ and $vec(\mathbf{F}^{*}\otimes\mathbf{F})=\mathbf{T}_{f}vec(\mathbf{f}\mathbf{f}^{H})$,  where $\mathbf{T}_f\in{\mathbb{C}}^{(N_2^{2}N_3^{2}){\times}(N_2^{2}N_3^{2})}$ is the transformation matrix formed by ones and zeros, which can be built by observing the relationship between $vec(\mathbf{F}^{*}\otimes\mathbf{F})$ and $vec(\mathbf{f}\mathbf{f}^{H})$. Then, (\ref{eq:15}) can be transformed into
\begin{eqnarray}
\nonumber
&&\|\mathbf{X}_{1}\mathbf{F}\mathbf{X}_{2}\mathbf{F}^{H}\mathbf{X}_{3}\|\\
 &=&\left\|\left(vec(\mathbf{X}_{2})^{T}\otimes(\mathbf{X}_{3}^{T}\otimes\mathbf{X}_{1})\right)\mathbf{T}_{f}vec(\mathbf{f}\mathbf{f}^{H})\right\|.
\end{eqnarray}

Inserting (\ref{eq:5}) into the numerator of eavesdropper's SNR (\ref{eq:8}) and omitting the terms involving second order channel uncertainties, the upper bound of ${\textmd{SNR}}_e$'s  numerator can be written as
\begin{eqnarray}\label{eq:16}
\nonumber
&&\quad\mathbf{g}_e\mathbf{W}\mathbf{H}\mathbf{q}\mathbf{q}^{H}\mathbf{H}^{H}\mathbf{W}^{H}\mathbf{g}_e^{H}\\\nonumber
&=&\mathbf{\bar{g}}_e\mathbf{W}\mathbf{H}\mathbf{q}\mathbf{q}^{H}\mathbf{H}^{H}\mathbf{W}^{H}\mathbf{\bar{g}}_e^{H}+2\mathfrak{Re}\{\triangle\mathbf{g}_e\mathbf{W}\mathbf{H}\mathbf{q}\mathbf{q}^{H}\mathbf{H}^{H}\mathbf{W}^{H}\mathbf{\bar{g}}_e^{H}\}\\\nonumber
&\leq&\mathbf{\bar{g}}_e\mathbf{W}\mathbf{H}\mathbf{q}\mathbf{q}^{H}\mathbf{H}^{H}\mathbf{W}^{H}\mathbf{\bar{g}}_e^{H}+2\varepsilon\|\mathbf{W}\mathbf{H}\mathbf{q}\mathbf{q}^{H}\mathbf{H}^{H}\mathbf{W}^{H}\mathbf{\bar{g}}_e^{H}\|\\\nonumber
&=&Tr\big(\mathbf{Z}\big((\mathbf{\bar{g}}_e^{H}\mathbf{\bar{g}}_e)^{T}\otimes(\mathbf{H}\mathbf{Q}\mathbf{H}^{H})\big)\big)+2\varepsilon\big\|\big(vec(\mathbf{H}\mathbf{Q}\mathbf{H}^{H})^{T}\\&&\otimes(\mathbf{\bar{g}}_e^{*}\otimes\mathbf{I}_M)\big)\mathbf{T}_{f}vec(\mathbf{Z})\big\|,
\end{eqnarray}
where the inequality holds by using (\ref{eq:14a}).
Similarly, the lower bound of ${\textmd{SNR}}_e$'s denominator can be expressed as
\begin{eqnarray}\label{eq:17}
\nonumber
&&\sigma_r^{2}\mathbf{g}_e\mathbf{W}\mathbf{W}^{H}\mathbf{g}_e^{H}+\sigma_e^2\\\nonumber
&=&\sigma_r^{2}\mathbf{\bar{g}}_e\mathbf{W}\mathbf{W}^{H}\mathbf{\bar{g}}_e^{H}+2\mathfrak{Re}\{\triangle\mathbf{g}_e\mathbf{W}\mathbf{W}^{H}\mathbf{\bar{g}}_e^{H}\}+\sigma_e^2\\\nonumber
&\geq&\sigma_r^{2}\mathbf{\bar{g}}_e\mathbf{W}\mathbf{W}^{H}\mathbf{\bar{g}}_e^{H}-2\varepsilon\|\mathbf{W}\mathbf{W}^{H}\mathbf{\bar{g}}_e^{H}\|+\sigma_e^2\\\nonumber
&=&Tr\big(\mathbf{Z}\big((\mathbf{\bar{g}}_e^{H}\mathbf{\bar{g}}_e)^{T}\otimes(\sigma_r^{2}\mathbf{I}_M)\big)\big)-2\varepsilon\big\|\big(vec(\mathbf{I}_M)^{T}\otimes(\mathbf{\bar{g}}_e^{*}\otimes\\&&\mathbf{I}_M)\big)\mathbf{T}_{f}vec(\mathbf{Z})\big\|,
\end{eqnarray}
where the inequality holds by using (\ref{eq:14b}).
Thus, the optimization problem (\ref{eq:9}) can be reformulated as
\begin{small}
 \begin{eqnarray}\label{eq:18}
\nonumber
 &&\mathop {\textrm{min}}_{\mathbf{Q},\mathbf{Z}}{\quad}Tr(\mathbf{Q})+Tr\big(\mathbf{Z}\big(\mathbf{I}_M\otimes\big(\mathbf{H}\mathbf{Q}\mathbf{H}^{H}+\sigma_r^{2}\mathbf{I}_M\big)\big),\\\nonumber
 &&s.t.{\quad}Tr(\mathbf{Z}\mathbf{A})\geq{r}_{th}^{(b)}\sigma_b^2,\\\nonumber
  &&{\qquad}Tr(\mathbf{Z}\mathbf{B}){\geq}2\varepsilon\big\|\big(vec(\mathbf{H}\mathbf{Q}\mathbf{H}^{H})^{T}\otimes(\mathbf{\bar{g}}_e^{*}\otimes\mathbf{I}_M)\big)\mathbf{T}_{f}\\\nonumber
  &&vec(\mathbf{Z})\big\|
+2{r}_{th}^{(e)}\varepsilon\big\|\big(vec(\mathbf{I}_M)^{T}\otimes(\mathbf{\bar{g}}_e^{*}\otimes\mathbf{I}_M)\big)\mathbf{T}_{f}vec(\mathbf{Z})\big\|,
\\
&&{\qquad}rank(\mathbf{Q})=1,rank(\mathbf{Z})=1,
 \end{eqnarray}
 \end{small}
 where
 \begin{small}
 \begin{eqnarray}
 \label{eq:19}&&\mathbf{A}=\big((\mathbf{g}_b^{H}\mathbf{g}_b)^{T}\otimes(\mathbf{H}\mathbf{Q}\mathbf{H}^{H})\big)-{r}_{th}^{(b)}\big((\mathbf{g}_b^{H}\mathbf{g}_b)^{T}\otimes(\sigma_r^{2}\mathbf{I}_M)\big),\\
 \label{eq:20}&&\mathbf{B}={r}_{th}^{(e)}\big((\mathbf{\bar{g}}_e^{H}\mathbf{\bar{g}}_e)^{T}\otimes(\sigma_r^{2}\mathbf{I}_M)\big)-\big((\mathbf{\bar{g}}_e^{H}\mathbf{\bar{g}}_e)^{T}\otimes(\mathbf{H}\mathbf{Q}\mathbf{H}^{H})\big).\quad\;\;\;
 \end{eqnarray}
 \end{small}
However, the optimizing problem (\ref{eq:18}) is non-convex due to the rank constraints. Therefore, we resort to semidefinite relaxation technique that firstly neglects these rank constraints, and the optimization problem turns to be
\begin{small}
\begin{subequations}\label{eq:23}
\begin{eqnarray}
 &&\mathop {\textrm{min}}_{\mathbf{Q},\mathbf{Z}}{\quad}Tr(\mathbf{Q})+Tr\big(\mathbf{Z}\big(\mathbf{I}_M\otimes\big(\mathbf{H}\mathbf{Q}\mathbf{H}^{H}+\sigma_r^{2}\mathbf{I}_M\big)\big),\\
 &&s.t.{\quad}Tr(\mathbf{Z}\mathbf{A})\geq{r}_{th}^{(b)}\sigma_b^2,\\\nonumber
  &&{\qquad}Tr(\mathbf{Z}\mathbf{B}){\geq}2\varepsilon\big\|\big(vec(\mathbf{H}\mathbf{Q}\mathbf{H}^{H})^{T}\otimes(\mathbf{\bar{g}}_e^{*}\otimes\mathbf{I}_M)\big)\mathbf{T}_{f}\quad\\
  &&vec(\mathbf{Z})\big\|
+2{r}_{th}^{(e)}\varepsilon\big\|\big(vec(\mathbf{I}_M)^{T}\otimes(\mathbf{\bar{g}}_e^{*}\otimes\mathbf{I}_M)\big)\mathbf{T}_{f}vec(\mathbf{Z})\big\|.\quad\;\;\;\;
 \end{eqnarray}
 \end{subequations}
\end{small}
Additionally, the above problem is still non-convex for both $\mathbf{Q}$ and $\mathbf{Z}$ due to the bilinear properties~\cite{paper:9}. Nevertheless, with fixed $\mathbf{Z}$, the problem is convex for $\mathbf{Q}$. Similarly, with fixed $\mathbf{Q}$, the problem is convex for $\mathbf{Z}$. Therefore, we can use iterative algorithm to solve the optimization problem (24), which is stated in Algorithm.~1.
\begin{algorithm}
\caption{Joint beamforming design of source and relay.} \label{algo_minballd}
\begin{algorithmic}[1]
\STATE Initialization:\\
Initialize the matrix $\mathbf{Q}^{(0)}=\frac{1}{N}P_{s}$, $\xi^{(0)}=10^3$, $\epsilon=10^{-3}$, $n=1$, ${N_{max}}=30$.
\STATE Iteration:\\
a) Compute $\mathbf{Z}^{(n)}$  by solving the problem (24) with fixed values of $\mathbf{Q}^{(n-1)}$.\\
b) Compute $\mathbf{Q}^{(n)}$ by solving the problem (24) with fixed value of  $\mathbf{Z}^{(n)}$. \\
c) Record the power soluton of problem (24) as  $\xi^{(n)}$.
\STATE Termination:\\
The algorithm terminates either when $\xi^{(n)}$ converges, i.e., $\mid\frac{\xi^{(n)}-\xi^{(n-1)}}{\xi^{(n)}}\mid\leq{\epsilon}$, or when $n\geq{N_{max}}$, where $\epsilon$ is a predefined threshold and $N_{max}$ is the maximum iteration number.\\
Output $\mathbf{Z}^{opt}=\mathbf{Z}^{(n)}$, $\mathbf{Q}^{opt}=\mathbf{Q}^{(n)}$.\\
Else, $n=n+1$, and go to step 2.
\end{algorithmic}
\end{algorithm}
To solve problem (\ref{eq:23}) we used CVX, a package for specifying and solving convex programs~\cite{paper:10}. Let us denote $\mathbf{Q}^{opt}$ and $\mathbf{Z}^{opt}$ as the solution obtained from $\mathrm{CVX}$. If $rank(\mathbf{Q}^{opt})=1$ and $rank(\mathbf{Z}^{opt})=1$ , then we can use eigenvalue decomposition to obtain the optimal $\mathbf{q}^{opt}$ and $\mathbf{w}^{opt}$; Otherwise, randomization technique can be applied to obtain $\mathbf{q}^{opt}$ and $\mathbf{w}^{opt}$~\cite{paper:12}. Specifically, we generate a set of random dual vectors which conform the Gaussian distribution, i.e.,  $\mathbf{\tilde{q}}\sim{\mathcal{N}(0,\mathbf{Q}^{opt})}$
 and $\mathbf{\tilde{w}}\sim{\mathcal{N}(0,\mathbf{Z}^{opt})}$. Among these dual vectors, there might exist the pairs that violate the constraints of (24). Accordingly, we use $\alpha$ and $\beta$ as the scale factors and denote $\mathbf{\hat{w}}=\alpha\mathbf{\tilde{w}}$ and $\mathbf{\hat{q}}=\beta\mathbf{\tilde{q}}$ as the new candidate pair. The values of $\alpha$ and $\beta$ could be obtained by setting the the constraints of (24) to equalities as shown in (25).
\newcounter{mytempeqncnt}
\begin{figure*}[t]
\normalsize
\setcounter{mytempeqncnt}{\value{equation}}
\setcounter{equation}{24}
\begin{eqnarray}\label{eq:25}
\alpha=\sqrt{\left(\frac{{r}_{th}^{(b)}\sigma_b^2}{Tr(\mathbf{\tilde{w}}\mathbf{\tilde{w}}^H\mathbf{A})}\right)}, \beta=\sqrt{\left(\frac{Tr(\alpha^2\mathbf{\tilde{w}}\mathbf{\tilde{w}}^H\mathbf{B})-2{r}_{th}^{(e)}\varepsilon\big\|\big(vec(\mathbf{I}_M)^{T}\otimes(\mathbf{\bar{g}}_e^{*}\otimes\mathbf{I}_M)\big)\mathbf{T}_{f}vec(\alpha^2\mathbf{\tilde{w}}\mathbf{\tilde{w}}^H)\big\|}
{2\varepsilon\big\|\big(vec(\mathbf{H}\mathbf{\tilde{q}}\mathbf{\tilde{q}}^H\mathbf{H}^{H})^{T}\otimes(\mathbf{\bar{g}}_e^{*}\otimes\mathbf{I}_M)\big)\mathbf{T}_{f}vec(\alpha^2\mathbf{\tilde{w}}\mathbf{\tilde{w}}^H)\big\|}\right)}
\end{eqnarray}

\setcounter{equation}{25}
\hrulefill
\vspace*{4pt}
\end{figure*}
Finally, the candidate pair that can achieve the minimum value of objective function (24a) can be viewed as a quasi-optimal solution. The randomization technique applied in this letter is summarized in Algorithm.~2.
\begin{algorithm}
\caption{Randomization technique for obtaining the source and relay precoders.} \label{algo_minballd}
\begin{algorithmic}[1]
\STATE Initialization:\\
Generate a set of $K$ random pairs of dual vectors $[\mathbf{\tilde{q}}^{(k)},\;\mathbf{\tilde{w}}^{(k)}]$ which conform the Gaussian distribution $\mathbf{\tilde{q}}^{(k)}\sim{\mathcal{N}(0,\mathbf{Q}^{opt})}$.
 and $\mathbf{\tilde{w}}^{(k)}\sim{\mathcal{N}(0,\mathbf{Z}^{opt})}$, $k=1,2,...,K$. Set $i$=0.
\STATE Computation:\\
a) $i=i+1$.\\
b) If the $i$-th pair $[\mathbf{\tilde{q}}^{(i)},\;\mathbf{\tilde{w}}^{(i)}]$ does not violate the constraints of (24), then we compute (24a) and record the value as $\textmd{OPT}_{value}^{(i)}$.\\
c) Otherwise, we compute the values of $\alpha$ and $\beta$ by using (25), and  compute $\mathbf{\hat{w}}=\alpha\mathbf{\tilde{w}}$ and $\mathbf{\hat{q}}=\beta\mathbf{\tilde{q}}$. Then, we use $[\mathbf{\hat{q}}^{(i)},\;\mathbf{\hat{w}}^{(i)}]$ as the new candidate pair to calculate (24a) and record the value as $\textmd{OPT}_{value}^{(i)}$.\\
d) If $i\neq{K}$, go to sub-step a). \\
\STATE Output:\\
Among all the values of $\textmd{OPT}_{value}^{(i)}, i=1,2,...,K$, we choose the smallest one and output its corresponding candidate pair vectors as the quasi-optimal solutions.
\end{algorithmic}
\end{algorithm}
\section{Simulation Results}
Numerical results are demonstrated in this section so as to verify the effectiveness of our proposed method. Without loss of generality, we set $\sigma_r^2=\sigma_b^2=\sigma_e^2=1$ and $M=N=4$. The simulation results are averaged over 1000 channel realizations.

Firstly, we investigate the power consumption versus different thresholds of (24) in Fig.~2. The non-robust precoding scheme corresponds to the case of setting $\varepsilon=0$ in (24). From Fig.~2, we can observe that with fixed $r_{th}^{(e)}$ and $r_{th}^{(b)}$, the robust precoding scheme will always consume more power than the non-robust precoding scheme, which is reasonable since the worst-case is considered in our robust scheme. Similar performance can also be seen in~\cite{paper:6}. Besides, for both of the robust beamforming scheme and the non-robust beamforming scheme, as the thresholds become tighter, more power comsumption is expected which is in consistent with our analysis. However, such comparison cannot show the actual performance of robust precoding scheme. The actual performance will be illustrated in Fig.~3.

Then, we examine distribution of the eavesdropper's SNR with distinct values of $\varepsilon$ and $r_{th}^{(e)}$. With fixed values of $\varepsilon$ and $r_{th}^{(e)}$, we can observe that for the non-robust precoding scheme almost half of eavesdropper's SNRs will be larger than the preset thresholds. Oppositely, the majority of our robust scheme's SNRs will be less than these thresholds. Additionally, since our designed beamforming vector is to constrain SNR of eavesdropper for the worst-case channel error, it might result in performance degradation for other channel error cases. Thus, that is why there are still SNRs that are larger than the thresholds for our robust precoding.
\begin{figure}[!htbp]
\centering
  \includegraphics[width=8cm,height=6cm]{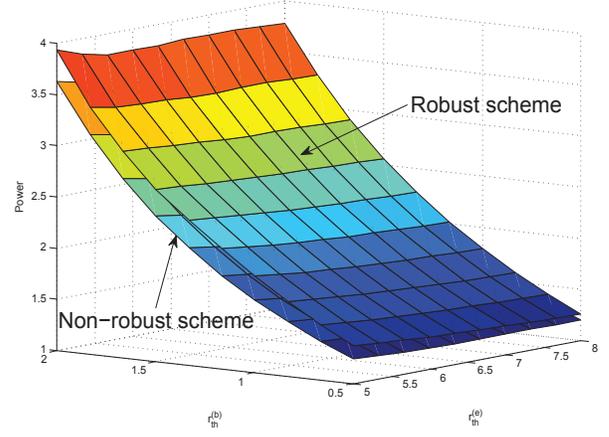}
\caption{Power consumption versus distinct values of $r_{th}^{(e)}$ and $r_{th}^{(b)}$, $\varepsilon=0.01$}
\label{fig_network_model}
\end{figure}
\begin{figure}[!htbp]
\centering
  \includegraphics[width=8.5cm,height=6cm]{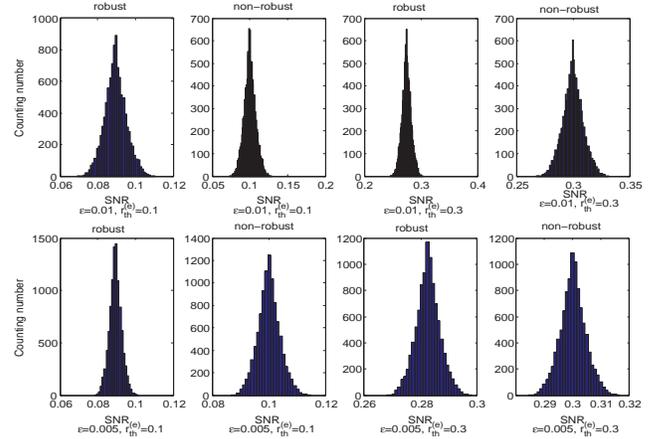}
\caption{Distribution of eavesdropper's SNR with distinct values of $\varepsilon$ and $r_{th}^{(e)}$ }
\label{fig_network_model}
\end{figure}
\section{Conclusion}
This letter proposes a source and relay secure optimization design with presence of channel uncertainty. It aims at minimizing the sum power consumption of source and relay while satisfying certain prefixed QoS requirements. Finally, simulation results verify the effectiveness of our algorithm compared with non-robust precoding scheme.


\begin{thebibliography}{1}
\bibitem{paper:1}
A. D. Wyner, "The Wire-Tap Channel," \emph{Bell Sys. Tech. J.}, pp. 1355-1387, October 1975.
\bibitem{paper:2}
S. Goel and R. Negi, "Guaranteeing Secrecy using Artificial Noise," \emph{Wireless Communications, IEEE Transactions on} , vol. 7, no. 6, pp. 2180-2189, June 2008.
\bibitem{paper:3}
C. Jeong, I. Kim and D. Kim, "Joint Secure Beamforming Design at the Source and the Relay for an Amplify-and-Forward MIMO Untrusted Relay System," \emph{Signal Processing, IEEE Transactions on}, vol. 60, no. 1, pp. 310-325, January 2012.
\bibitem{paper:4}
W. Liao, T. Chang, W. Ma and C. Chi, "QoS-Based Transmit Beamforming in the Presence of Eavesdroppers: An Optimized Artificial-Noise-Aided Approach," \emph{Signal Processing, IEEE Transactions on}, vol. 59, no. 3, pp. 1202-1216, March 2011.
\bibitem{paper:5}
A. Tajer, N. Prasad, and X. Wang, "Robust linear precoder design for multi-cell downlink transmission," \emph{Signal Processing, IEEE Transactions on}, vol. 59, no. 1, pp. 235-251, 2011.
\bibitem{paper:6}
B. K. Chalise and L. Vandendorpe, "MIMO Relay Design for Multipoint-to-Multipoint Communications With Imperfect Channel State Information," \emph{Signal Processing, IEEE Transactions on}, vol. 57, no. 7,
pp. 2785-2796, July 2009.
\bibitem{paper:7}
Y. Pei, Y. Liang, L. Zhang, K. Teh and K. Li, "Secure communication over MISO cognitive radio channels," \emph{Wireless Communications, IEEE Transactions on}, vol. 9, no. 4, pp. 1494-1502, April 2010.
\bibitem{paper:8}
R. A. Horn and C. R. Johnson, \emph{Matrix Analysis}, Cambridge University Press, 1985.
\bibitem{paper:9}
S. Boyd and L. Vandenberghe, \emph{Convex Optimization}, Cambridge University Press, 2004.
\bibitem{paper:10}
CVX Research, Inc. CVX: Matlab software for disciplined convex programming, version 2.0 beta. http://cvxr.com/cvx, September 2012.
\bibitem{paper:12}
Y. Huang and D. P. Palomar, "Rank-constrained separable semidefinite programming with applications to optimal beamforming," \emph{Signal Processing, IEEE Transactions on}, vol. 58, no. 2, pp. 664-678, February 2010.
\end{thebibliography}
\end{document}